\documentclass[12pt,preprint]{aastex}

\newcommand{\myemail}{lic@nju.edu.cn}

\shorttitle{Particle acceleration during GLE71}
\shortauthors{Li et al.}

\begin{document}

\title{Electron and Proton Acceleration during the First GLE Event of the Solar Cycle 24}

\author{C. Li\altaffilmark{1,2}, Kazi A. Firoz\altaffilmark{3}, L. P. Sun\altaffilmark{1}, and L. I. Miroshnichenko\altaffilmark{4,5}}

\altaffiltext{1}{School of Astronomy \& Space Science, Nanjing University, Nanjing 210093, China.\\ \myemail}
\altaffiltext{2}{Key Laboratory for Modern Astronomy and Astrophysics (Nanjing University), Ministry of Education, Nanjing 210093, China}
\altaffiltext{3}{Key Laboratory of Dark Matter \& Space Astronomy, Purple Mountain Observatory, Chinese Academy of Science, Nanjing 210008, China}
\altaffiltext{4}{N. V. Pushkov Institute of Terrestrial Magnetism, Ionosphere and Radio Wave Propagation (IZMIRAN), Russian Academy of Sciences, Troitsk, 142190 Moscow Region, Russia}
\altaffiltext{5}{D. V. Skobeltsyn Institute of Nuclear Physics (SINP), M. V. Lomonosov Moscow State University, 1(2), Vorobyevy Gory, 119234 Moscow, Russia}

\begin{abstract}
High-energy particles were recorded by the near-Earth spacecraft and ground-based neutron monitors (NMs) on 2012 May 17. This event was the first Ground Level Enhancement (GLE) of the solar cycle 24. In present study, we try to identify the acceleration source(s) of solar energetic particles (SEPs) by combining \emph{in-situ} particle measurements from $WIND$/3DP, $GOES$ 13, and solar cosmic rays (SCRs) registered by several NMs, as well as the remote-sensing solar observations from $SDO$/AIA, $SOHO$/LASCO, and $RHESSI$. We derive the interplanetary magnetic field (IMF) path length (1.25 $\pm$ 0.05 AU) and solar particle release (SPR) time (01:29 $\pm$ 00:01 UT) of the first arriving electrons by using their velocity dispersion and taking into account the contamination effects. It is found that the electron impulsive injection phase, indicated by the dramatic change of spectral index, is consistent with the flare non-thermal emission and type III radio bursts. Based on the potential field source surface (PFSS) concept, a modeling of the open-field lines rooted in the active region (AR) has been performed to provide escaping channels for flare-accelerated electrons. Meanwhile, relativistic protons are found to be released $\sim$10 min later than the electrons, assuming their scatter-free travel along the same IMF path length. Combining multi-wavelength imaging data on the prominence eruption and coronal mass ejection (CME), we obtain some evidence of that GLE protons, with estimated kinetic energy of $\sim$1.12 GeV, are probably accelerated by the CME-driven shock when it travels to $\sim$3.07 solar radii. The time-of-maximum (TOM) spectrum of protons is typical for the shock wave acceleration.
\end{abstract}


\section{Introduction}

Large solar energetic particle (SEP) events draw more and more research enthusiasm not only because of theoretical interest to high-energy solar phenomena but also due to their perspectives for space weather forecasting. The relativistic extension of some large SEP events can produce sufficient secondary particles registered by ground-based neutron monitors (NMs). As a result, we observe so-called Ground Level Enhancement (GLE) of solar cosmic rays (SCRs). To understand where and how solar particles are accelerated to high energies in large SEP events is one of the main topics of space physics. This issue, however, still remains controversial \citep{mir08,rea09}.

Historically, it was thought for many years that the flares are main sources of GLEs \citep[e.g.][]{mir01}. The modern paradigm \citep{kah94,kah01,rea99,rea02,rea09,cli06,gop12} suggests that in large SEP events, especially in GLE events, particle acceleration mainly takes place at the shocks driven by coronal mass ejections (CMEs) rather than in flare active regions (ARs). However, large SEP events are always associated with flares and CMEs concomitantly (without exception in GLE events), both of them being the different manifestations of the same process of magnetic energy release \citep{har95,zha01,wan03}. Theoretically, the flares \citep[e.g.][]{som11} and CME-driven shocks \citep[e.g.][]{zan00,ber03} both are capable of accelerating charged particles to high energies. Observationally, we also have certain evidence that particle acceleration directly in flare sites cannot be ruled out \citep{can02,can06,can10}. The question arises: what dominates SEP injection?

According to recent observations and modeling of the GLEs, the answer is mostly consistent with a flare-associated prompt component (PC) and a CME-associated delayed component, DC \citep[e.g.,][]{li07a,li07b,li09,vas11,asc12}. Assuming existence of open-field lines originated in ARs, a direct flare contribution should exist in large SEP events. Using potential field source surface (PFSS) model, \citet{sch03} found that a significant fraction of the IMF lines are directly connected to magnetic plage of ARs. These lines may provide escaping channels for particles accelerated at low coronal sites. By applying the same model, a statistical link between the coronal magnetic topologies and dynamics of \emph{in-situ} electrons was recently established \citep{li10}. For individual SEP event, a more practical and accurate method to construct coronal magnetic fields is absolutely necessary \citep{li11}.

Timing of SEPs with respect to flare emission (e.g., hard X-ray production, type III radio bursts) and CME signatures (e.g., type II radio bursts) is generally a tool for identification of particles acceleration source. One of the first statistical studies of the solar particle release (SPR) times was performed by \citet{cli82} who found that $\sim$100 keV electrons were released within 5 min earlier than $\sim$2 GeV protons, followed by $\sim$MeV electrons at least 5 min later. \citet{kah03} studied the first 10 GLE events of solar cycle 23 and found that in half of the event number an injection of near-relativistic electrons was preceded by the GeV protons. More recently, \citet{sim06} carried out a comprehensively study of particle timing for the GLE of 2005 January 20. It was found that the injection of near-relativistic electrons is delayed by $\sim$6 min from that of the GeV protons. The author suggested that the protons were directly related to the flare; however, the CME was responsible for delaying release of the flare electrons onto IMF lines connected to the spacecraft. According to these studies, there seems to be no systematic trend for acceleration of different particle species in individual events. This may arise from several reasons, such as selective acceleration \citep{can86,mil97}, cross-field transport or perpendicular diffusion \citep{qin11,wan12}, particle trapping in coronal loops, coronal magnetic topology as mentioned above, etc.

The event under study in this paper occurred on 2012 May 17. From the beginning of the solar cycle 24 (2009 January) up to now, it was the first and single GLE event, whereas previous cycle 23 had produced five GLEs in the first 4.5 years. During the event, high-energy solar particles were recorded by both near-Earth spacecraft and some ground-based NMs, along with a medium-strength solar flare (1F/M5.1) and a high-speed CME (1582 km/s). It enables us to carry out a cross-disciplinary investigation, by combining \emph{in-situ} particle measurements and NM data, as well as the results of remote-sensing solar observations. The main goal of this study is to extend our understanding of particle acceleration in large SEP events.

First of all, we describe in detail particle measurements and methods of data analysis (Section 2) to investigate electrons and protons release at the Sun. The next Section 3 presents available solar observations, including the data on the SEP-associated flare and concomitant CME. Section 4 is devoted to discussion on possible solar source(s) of the two species, by systematically analyzing event timing, particle energy spectrum, etc. Summary and discussion are given in Section 5.

\section{Particle measurements}

\subsection{Release of electrons}

Spacecraft $WIND$ is currently orbiting satellite in the Sun-Earth L1 libration point, with the experiment on registration of non-thermal electrons with energies from a few keV up to near-relativistic energy range. The three-dimensional Plasma and Energetic Particles instrument \citep[3DP;][]{lin95} onboard the $WIND$ observes electrons with the solid-state telescopes (SSTs) from 27 to 517 keV with a time resolution of $\sim$12 sec. For SSTs, laboratory calibration shows that a proportion of incident electrons will scatter back out of each detector before fully depositing their original energy and produce secondary particles that contaminate lower energy channels. This leads to erroneous results when studying the electrons release from the Sun if the contamination is not taken into account. \citet{wan09} corrected empirically the count contamination by assuming that $\sim$10 -- 25$\%$ of incident electrons will be scattered out of each energy channel and these electrons will be evenly distributed over the lower energy channels.

To correct the count contamination, here we apply the correction matrix, $C$, that can be deduced, in principle, from the response function $g^{k}(E_{i})$ \citep{hag03} as
\begin{equation}
C_{k,n}=\int_{E_{n}^{d}}^{E_{n}^{u}}g^{k}(E_{i})dE_{i},
\end{equation}
where $E_{n}^{d}$ ($E_{n}^{u}$) is the upper (down) limit of the energy channel $n$. The corrected flux, $I$, is deduced from multiplication of the observed flux $i$ by the correction matrix as
\begin{equation}
I\left(
\begin{array}{c}
I_{1}\\I_{2}\\\vdots\\I_{n-1}\\I_{n}
\end{array}
\right)
=
C\left(
\begin{array}{ccccc}
1&C_{1,2}&\cdots&C_{1,n-1}&C_{1,n}\\
0&1&\cdots&C_{2,n-1}&C_{2,n}\\
\vdots&\vdots&\ddots&\vdots&\vdots\\
0&0&\cdots&1&C_{n-1,n}\\
0&0&\cdots&0&1
\end{array}
\right)
\otimes
i\left(
\begin{array}{c}
i_{1}\\i_{2}\\\vdots\\i_{n-1}\\i_{n}
\end{array}
\right).
\end{equation}
In the practical implementation, however, the coefficients of correction matrix are given empirically varying from 0.05 to 0.3 for different energy channels. The corrected flux can be then validated by comparing its velocity dispersion (onset times are later for lower energies) with an expected one \citep[see][]{sun12}.

Figure 1 shows the electron intensity profiles observed by $WIND$/3DP/SSTs in seven channels before (black lines) and after (color lines) correction on 2012 May 17. It is found that lower energy channels suffer heavier contaminations due to more scattered particles from higher energy channels. Another fact is that the correction is more effective at the particle injection phase, resulting in an expected clear velocity dispersion.

A linear fit to the velocity dispersion has been commonly used to study solar particle release (SPR) times of beam-like SEP events \citep{lin81,rea85,kru99}. Recently, \citet{rea09} successfully applied this method to investigate the SPR times for protons and ions of He, O, and Fe during the 16 GLE events that occurred from 1994 to 2007, in solar cycle 23. Note that the author did not include electron data in his consideration. Assuming the first arriving particles are traveling along the IMF path length without scattering, the SPR time of electrons with energy $E_{n}$ can be expressed as
\begin{equation}
T_{SPR}(E_{n})=T_{onset}(E_{n})-L/v(E_{n}),
\end{equation}
where $T_{onset}(E_{n})$ is the onset time of electron flux increase at 1 AU for each energy channel ($n=1,2,\cdots7$), $L$ is the IMF path length from electrons release site on the Sun to the spacecraft, and $v(E_{n})$ is the velocity of electrons. Note that $T_{onset}(E_{n})$ is determined by the first time when a background subtracted flux exceed the level of 3$\sigma$ (standard deviation), and the error bars are determined by times of $\pm 3\sigma$ excess around $T_{onset}(E_{n})$.

We apply Equation (3) to the $WIND$/3DP/SSTs corrected electron data (color lines of Figure 1). Plotting onset times versus $v^{-1}$ yields a line with the initial SPR time as the intercept and the slope corresponds to the IMF path length, which is shown in Figure 2. The electron SPR time in these terms is 01:29 $\pm$ 00:01 UT, and the IMF path length is 1.25 $\pm$ 0.05 AU. Note that the SPR time of protons (see Section 2.2) is also marked by triangle on the time axis.

To further confirm this evaluation, assuming charged particles travel along the spiral IMF lines, the path length can be calculated by solution of the IMF equation deduced from the solar wind model \citep{par58}. In a polar coordinate system, the IMF equation is expressed as
\begin{equation}
\varphi=\frac{\omega}{u}(r_{0}-r),
\end{equation}
where $\varphi$ is the azimuth angle of the IMF footpoint on the solar surface, $\omega$ is the angular speed of solar rotation, $u$ is the solar wind speed, and $r$ is the radial distance from the Sun center, $r_{0}=1AU$ at the near-Earth space. Therefore, combing Equation (4), a formula for the estimation of nominal length $L$ of IMF line may be given as
\begin{equation}
dL=\sqrt{(dr)^{2}+(rd\varphi)^{2}}=\sqrt{1+(\frac{\omega}{u}r)^{2}}dr.
\end{equation}

Integrating the Equation (5) by $r$ from the solar radius $R_{s}$ to the Earth's orbit $r_{0}$, adapting the angular speed of the solar rotation $\sim$1.7$\times10^{-4}$ deg/s and taking into account the solar wind speed $\sim$360 km/s before this event, we get finally that the value of $L$ to be $\sim$1.21 AU. Then taking $v(E_{7})$ to be approximately 0.9$c$ (where $c$ is the velocity of light) corresponding to the highest energy channel $E_{7} = 517$ keV, the electron SPR time derived from Equation (3) is to be $\sim$01:30 UT. Therefore, this estimate is quite consistent with the previous one.

\subsection{Release of protons}

To study the SPR time of protons, we analyze proton data in the energy range of $\sim$31 -- 433 MeV obtained onboard the $Geostationary\ Operational\ Environment\ Satellite\ (GOES)$ 13 with a time resolution of 1 min (top panel of Figure 3). The averaged 1-min intensities of SCRs observed at 4 NMs (bottom panel of Figure 3) were also used from the Neutron Monitor Database (NMDB; $\url{http://www.nmdb.eu}$), which collects data from many NMs that operate permanently at different sites around the globe (the worldwide network of cosmic ray stations). In total, for the GLE71 event there are records of 17 NMs from about 50 stations \citep{kle12}.

The GLE events characterize only relativistic part of entire energy spectrum of SCRs (kinetic energy $E_{p} \geq$ 433 MeV/nucleon, or magnetic rigidity $R \geq$ 1 GV). If the energy of primary protons is $E_{p} <$ 100 MeV ($R <$ 0.44 GV) neutron monitors are practically do not respond them due to atmospheric absorption of neutrons (so-called ``atmospheric cutoff", $R_{a}$). A maximum of the NM response is within 1 -- 5 GV. It means that all high-latitude (polar) NM stations start to record secondary neutrons efficiently from the same rigidity of the primary protons about 1 GV (433 MeV), irrespective of the NM nominal ``geomagnetic cutoff rigidity", $R_{c}$. As it fortunately happened, rigidity $R \sim$ 1.0 GV ($E_{p} \sim$ 433 MeV) is approximately midway between the low rigidity and ultra-relativistic rigidity range, and it turned out to be a convenient reference point as a characteristic rigidity cutoff at the polar NM stations \citep{sma96}.

As seen from the top panel of Figure 3, the $GOES$ 13 has recorded a fast rise in the flux of non-relativistic solar protons, followed by a slower decay, which was still ongoing on 2012 May 18. Note that several non-relativistic proton events stronger than that of 2012 May 17 were detected by $GOES$ in 2012 January and March. But the event of 2012 May 17 was obviously extended in the range of much higher energies than the other ones, being considerably weaker at lower energies. Through our research (see Section 4.3) we try to understand the reason for these differences. Whereas, relativistic SCR increases have been mostly recorded by NMs at high (polar) latitudes, at geomagnetic cutoff rigidities $R_{c} <$ 1 GV. Bottom panel of Figure 3 presents four the most intensive increases of the GLE71 recorded at SOPB (South Pole Bare, $R_{c} =$ 0.1 GV), SOPO (South Pole, 0.1 GV), APTY (Apatity, 0.65 GV), and OULU (Oulu, 0.8 GV).

As it was found by the data of four NMs, the first relativistic solar protons started to arrive to the Earth at nearly the same time $\sim$01:51 UT, in spite of that two of NMs are located in the southern hemisphere and two others -- in the northern hemisphere. As one of the possibilities, it may be evidence of isotropic flux of SCRs. On the other hand, according to \citet{kle12}, some SCR increases have been also recorded by five other NMs (Kerguelen, Yakutsk, Newkirk, Magadan, and Kiel) with $R_{c}$ above 1 GV (1.14, 1.65, 2.10, 2.10, and 2.36 GV, respectively). The authors also believe that no signal was observed at $>$3 GV cutoff. In total, by surface observations, there are 5-minute data of 17 NMs. However, the signals of the most of them are at the level of statistical fluctuations, so there was, in practice, no possibility to build up a curve of latitude effect in cosmic ray intensity. In fact, no latitude effect has been registered by NMs at the Earth's surface that turned out to be a simple consequence of rather soft spectrum of accelerated protons (but not the result of SCR anisotropy). This may suggest that the SEPs on 2012 May 17, in general, and GLE71 event had, indeed, rather soft energy spectrum (see Section 4.3).

Based on Figure 3, we can estimate SPR time of protons applying Equation (3), under the assumption that the first arriving protons traveled along the same IMF path length, 1.25 $\pm$ 0.05 AU, as the electrons did. If we take $v$ to be approximately 0.7$c$ corresponding to the highest energy channel of 433 MeV for $GOES$ non-relativistic protons, then the evaluated SPR time is 01:40 $\pm$ 00:03 UT. For GLE relativistic protons with energy of $\geq$GeV, we take $v$ to be approximately 0.9$c$, then the evaluated SPR time is 01:39 $\pm$ 00:02 UT (see also Figure 2). Considering the systematic errors, we believe that the GLE protons belong to the same population and form the relativistic extension of the first arriving $GOES$ protons.

From the above analysis, it also follows that the SPR time of GLE protons is $\sim$10 min later than the near-relativistic electrons (01:29 $\pm$ 00:01 UT). Assuming scatter-free propagation in interplanetary space, we suggest that the most probable reason of such discrepancy may rise from different acceleration source(s) of the two particle species.

\section{Solar observations}

\subsection{Flare active region}

To identify the acceleration source(s) of SEPs, we first study the coronal morphology and magnetic topology of the AR 11476. Left panel of Figure 4 shows the magnetic field lines derived from the potential-field source-surface (PFSS) model \citep{sch03} and overlaid on the magnetogram obtained from the Helioseismic and Magnetic Imager \citep[HMI;][]{sch12}. Right panel of Figure 4 presents the two-ribbon flare structure obtained from the Atmospheric Imaging Assembly \citep[AIA;][]{lem11} on board the $Solar\ Dynamics\ Observatory\ (SDO)$ contoured with the $RHESSI$ HXR sources.

One can see that the closed field lines straddle on the two flare ribbons (see in 1600 $\rm {\AA}$ image, right panel of Figure 4) and show loop-like structures (see in 171 $\rm {\AA}$ images, left column of Figure 5 below). The open field lines rooted in the AR with negative polarity that can provide escaping channels for flare accelerated particles. The 12 -- 25 keV thermal bremsstrahlung source shows a loop-top structure, whereas the 50 -- 100 keV non-thermal sources just located at the ribbons, in other words, in the footpoints of the loop. This is consistent with the classical model of solar eruption, which introduces a loop-top HXR source and two sources at the footpoints on the ribbons.

According to the classical picture of a two-ribbon flare/three-component CME model, when a flux rope (filament on the solar surface or prominence on the solar limb) loses equilibrium and travels upward, an extensive reconnection current sheet (RCS) forms below the flux rope. As a result, inside the RCS region a great number of particles can be accelerated to extremely high energies. The upward motion drives the expansion of coronal loops above the flux rope to form the front of a CME, whereas the flux rope itself forms the core of the CME.

To display these consequences, we present temporal evolution of the AR observed in $SDO$/AIA 171 $\rm {\AA}$ and 304 $\rm {\AA}$ images shown in Figure 5. At $\sim$01:25 UT, a system of coronal loops is identified in 171 $\rm {\AA}$, beneath which a prominence is clearly seen in 304 $\rm {\AA}$. In fact, the prominence started to form much earlier, at $\sim$00:55 UT. At $\sim$01:34 UT, the prominence erupted and the coronal loops are stretched to ``open" out of the field of view (FOV). In reality, the prominence started to erupt as earlier as at $\sim$01:32 UT. It should be noticed that the time of prominence eruption is very close to the peak time of flare non-thermal emission and electron SPR time. At $\sim$01:55 UT, a thin line structure is identified at both wavelength, which indicates the trajectory of the prominence eruption, and such a structure may be a candidate of the RCS.

Based on the observations and modeling shown in Figure 4 and 5, we propose a sketch of coronal magnetic topology (see Figure 6) associated with the two-ribbon flare and prominence eruption. Particles accelerated in the RCS travel downward along closed field lines to produce HXR emission \emph{via} bremsstrahlung and flare ribbons \emph{via} collision with the chromosphere. A portion of high-energy particles can be scattered or transported by diffusion perpendicularly to the nearby open field lines and escape to interplanetary space.

On the other hand, particles can be selectively accelerated in the RCS, for instance, electrons have the priority to be accelerated than protons in the RCS via wave-particle interactions \citep{can86,mil97}. It explains why the electrons release at the Sun is consistent with the flare emission and type III radio bursts (see Section 4.1), while for the protons release this is not valid.

\subsection{CME and the associated shock}

Out FOV of $SDO$/AIA, the Large Angle and Spectrometric Coronagraph \citep[LASCO;][]{bru95} on board $SOHO$ provides white light (WL) observations of the high-speed CME. Figure 7 shows the LASCO C2 running difference image at $\sim$01:48 UT with a previous image subtracted. The three components of CME are identified as core (CO), cavity (CA), and leading front (LF). Combing Figure 4 and 5, it may be found that the trajectory of the prominence eruption is along the direction of the CME core, and the expansion of coronal loops forms the leading front of the CME. It is also found a diffusive structure or a signature of wave propagating ahead of the CME LF, which is probably the CME-driven shock \citep{vou03}.

If one refers to the CME catalog ($\url{http://cdaw.gsfc.nasa.gov/CME_{-}list/index.html}$), the CME sky-plane velocity from linear extrapolation in this case to be $\sim$1582 km/s, which has a minor projection effect since it originated on the solar limb. The liftoff time of the CME is extrapolated onto the solar surface at $\sim$01:32 UT that coincides with the onset time of the prominence eruption as it was already mentioned before.

We can then evaluate the CME height with respect to protons release. At the SPR time of GLE protons (01:39 UT + 8.3 min = 01:47.3 UT, 8.3 min is a light travel time of 1 AU.) the CME reaches $\sim$3.07 $Rs$. This is consistent with the conclusion by \citet{gop12} who studied the GLE events of solar cycle 23 and found the release of protons occurs when the CMEs reach an average height of $\sim$3.09 $Rs$ for well connected events with the source helio-longitudes in the range of W20 -- W90. It is timely to note that this result does not contradict to theoretical estimate of the radius of $\sim2.0 - 3.0 R_{s}$ for effective particle acceleration by spherical shock wave \citep{ber03}.

This CME and the associated coronal shock probably resulted in an interplanetary shock passage which was observed by the $ACE$ satellite at approximately 01:36 UT on 2012 May 20. The geomagnetic field responded with an isolated active period during 03:00 ¨C 06:00 UT on 2012 May 20.

\section{Relations to the acceleration source}

\subsection{Event timing}

Particle injection phase with respect to solar multi-wavelength emission is generally considered as a key to understand SEPs acceleration. Here, we demonstrate that electrons injection phase can be displayed in the form of spectral evolution, by applying a power-law spectrum $f(E_{e})\propto E_{e}^{-\gamma}$. The most dramatic change of spectral index $\gamma$, obviously, corresponds to the moment of impulsive injection phase. To do this, we use the electron SPR time (01:29 UT + 8.3 min = 01:37.3 UT) as a reference time with respect to solar multi-wavelength emission. The onset time obtained from intensity profile of every energy channel (see Figure 1) is shifted to the reference time. By fitting to the power-law spectrum, the evolution of electron spectral index is derived and shown in Figure 8 (top panel). The time range between two vertical lines indicate the proposed most impulsive phase of electrons injection, which is compared with solar multi-wavelength emission.

As shown in Figure 8, the 1F/M5.1 flare recorded in soft X-ray (SXR) range of 1 -- 8 $\rm {\AA}$ started at 01:25 UT, peaked at 01:47 UT, and followed by a $\sim$2 hr decay phase. At that time, the AR 11476 was situated at the position of N13W83 at the Sun, i.e. near the W-limb of the Sun. This position is roughly well-connected by IMF lines linking the Sun to near-Earth space. The time derivative of SXR flux (dashed line in positive values) is generally considered as a good approximation of the hard X-ray (HXR) flux according to the Neupert effect \citep{neu68}. It shows a similar profile with the microwave emission in 15.4 GHz, suggesting that the flare accelerated electrons are traveling downwards to generate HXR emission via bremsstrahlung or are trapping in magnetic loops to generate microwave emission via synchrotron. Whereas, the microwave emission in 2.7 GHz shows a few pulses in the decay phase, which are not clearly manifested at 15.4 GHz.

On the other hand, the flare accelerated electrons are traveling upwards along open field lines to generate type III radio bursts, and this was demonstrated by the radio dynamic spectra in the frequency range from 20 kHz to 14 MHz (Figure 8, two bottom panels). The type III radio bursts started at $\sim$01:33 UT, with a group of intensive emission lasting until to $\sim$01:44 UT. We found that the electron impulsive injection phase started at the peak time of HXR and microwave emission, at the most intensive emission of type III radio bursts, and lasted until to the decay phase of HXR and microwave emission.

The type II radio bursts, representing local electron acceleration at a shock wave moving out through corona, were reported to have the onset at $\sim$01:32 UT \citep{gop13}. This is consistent with the prominence eruption and CME liftoff. Note that the proton SPR time is $\sim$15 min later than the onset time of type II radio burst, suggesting that protons injection occurred at a higher corona site ($\sim$3.07 $Rs$) when the CME-driven shock develops to be an effective accelerator.

\subsection{Energy of GLE protons}

Another way to clarify the problem of solar sources for the GLE events is to check whether flare or CME is capable of producing relativistic protons. Theoretically, the answer is ``Yes" (see Section 1) for both proposed mechanisms of SEP acceleration (flare magnetic reconnection and CME-driven shock). Observationally, for individual GLE event, it is necessary to evaluate proton energy $E_{p}$ from its estimated time of travel to the Earth along the IMF path length under the assumption of scatter-free propagation. \citet{fi11a,fi11b} proposed empirical method to on the determination of possible relativistic energies of GLE69 and GLE70 that we testified recently for the interpretation of possible acceleration mechanisms for the same two events \citep{fir12}. Deducing from observations the time lag $\Delta t$ of GLE onset with respect to the related flare or CME, we can obtain the velocity of GLE protons $v_{p}=L/(\Delta t + 8.3min)$. Where $L$ is the IMF path length of protons from release site on the Sun to the Earth's orbit ($\sim$1.25 AU for this event). Then the energy of GLE protons may be roughly estimated from classic relativistic expression
\begin{equation}
E_{p}=(\gamma - 1)m_{p}c^{2}=(\frac{1}{\sqrt{1-(\frac{v_{p}}{c})^{2}}}-1)m_{p}c^{2},
\end{equation}
where $m_{p}c^{2}$ is the proton rest energy of 938.27 MeV.

Let us first assume that GLE protons are accelerated at the flare site. The time lag $\Delta t$ of GLE onset ($\sim$01:51 UT) with respect to the peak time of flare non-thermal emission ($\sim$01:36 UT) is $\sim$15 min. Based on Equation (6), the energy of protons is evaluated to be $\sim$0.11 GeV ($R \sim$ 0.45 GV), much less than the typical relativistic energy ($\sim$GeV) of the GLE protons and even less than the geomagnetic cut-off energy at polar NMs ($\sim$433 MeV, or $R \sim$ 1 GV). Therefore, the GLE protons are not likely caused by flare acceleration. Let us then assume GLE protons are accelerated at the CME-driven shock when it reaches at $\sim$3.07 $Rs$, where the shock developed to be an efficient accelerator. The time lag $\Delta t$ is $\sim$4 min, leading to the energy of protons to be $\sim$1.12 GeV ($R \sim$ 1.83 GV), which is sufficient to cause a GLE event.

In this context, it should be noted that according to theoretical model by \citet{zan00}, rather high energies may be reached at the early stages of shock evolution. In particular, energies of order of 1 GeV are possible for young shock waves. On the other hand, as it was convincingly shown by \citet{ber03}, with heliocentric distance $r$ the efficiency of acceleration by spherical shock wave decreases rather rapidly, and it causes the effective SCR acceleration seems to terminate when the shock reaches a distance of r $\sim$ 2 -- 3 $Rs$. As a result, SCR particles intensively escape from the vicinities of the shock.

\subsection{Time-of-maximum spectrum for protons}

As noted above (Section 2.2, Figure 3), the GLE71 turned out to be rather small, and its latitude effect, unfortunately, was not manifested in due magnitude. In reality, no latitude effect was registered by NMs at the Earth's surface. Therefore, a standard method of energy spectrum evaluation \citep{vas11} is not applicable in this case. Moreover, there were some suggestions that, in fact, the first GLE of current solar cycle have been already registered on 2012 January 27 -- 28. To make this point clearer, we apply the method of so-called time-of-maximum (TOM) spectrum \citep[see e.g.,][]{mir96,mir08}. As it was noted by those authors, the TOM spectrum of SEPs is a rough proxy of their source spectrum, at least, for well-connected events \citep[][]{mir73,for86}.

The spectrum of the January event is estimated by the $GOES$ 13 data obtained at energy thresholds of $E_{p} >$10, $>$50, and $>$100 MeV. According to our estimates by three intensity points, integral TOM spectrum of protons may be given in the form of $I(>E_{p}) \propto E_{p}^{-\alpha}$, where power-law index of the spectrum is $\alpha \sim$ 2.0 (the index for differential spectrum is $\gamma = \alpha + 1$). Another significant SEP event occurred on 2012 March 7. In the range of non-relativistic energies it may be considered as a largest one in the current solar cycle 24. Indeed, by $GOES$ 13 measurements, maximum proton intensity above 10 MeV reached about 6530 $pfu$. From the same data we estimate integral intensities of 305, and 70.5 $pfu$ for the protons $>$50, and $>$100 MeV, respectively. Again we get integral TOM spectrum with the index $\alpha \sim$ 2.0 ($\gamma \sim$ 3.0).




With this in mind, we also apply the TOM method to the event of 2012 May 17 to derive the integral spectrum of non-relativistic protons by using five energy channels ($E_{p} >$10, $>$30, $>$50, $>$60, and $>$100 MeV) of $GOES$ 13 data as shown in Figure 9 (left panel). Hence, we get $\alpha \sim$ 2.2 (for differential spectrum this value corresponds to $\gamma \sim$ 3.2). One can see from Figure 9 (right panel) that the power-law spectrum is broken over $\sim$100 MeV and displays an exponential high-energy tail (see below).

In this context, it is of great interest to estimate integral intensity of solar protons above 433 MeV by real NM data for this event. It may be done by the method of integral multiplicities of generation (or specific yield function) for neutron component developed by \citet{ile78} \citep[see also][]{mir01}. This method was elaborated for 1-hour data of neutron monitors. Averaging the data of Figure 3 (bottom panel) we get that a 1-hour amplitude of increase at Apatity (or Oulu) station was about 10$\%$. Hence, we get the value $I(>433$ MeV) $= 7.2\times10^{-4} pfu$. Unfortunately, our method \citep{ile78} provides intensity estimates only with accuracy factor of about 2.0. Nevertheless, we note that above estimate does not contradict to the observed integral intensity of protons with the energy $I(>700$ MeV)$= 2.3\times10^{-4} pfu$ (see left panel of Figure 9).

Table 1 gives a summary of our spectral estimates for all three SEP events of 2012, in comparison with that for well-known SEP event of 1972 August 4 (GLE24). Corresponding values of proton intensities for GLE24 were taken from our previous publications \citep{mir96,mir08}. From Table 1 it may be concluded that SEP events of 2012 January 27 and March 7, most likely, had no extension into relativistic range. On the other hand, between 2011 January -- 2012 May the PAMELA spectrometer registered several solar events with $>$100 MeV protons. The most powerful was event on 2012 March 7 \citep{baz12}. There are also some preliminary data on the effect of this SEP event registered by special huge muon telescope \citep{mak12} in El Leoncito (CASLEO, Argentina). Obviously, in the light of such observational indications, all NMDB data should be undergo to thorough additional analysis. We believe that in the course of this analysis a tendency of SEP spectra to steepening with energy increase \citep{mir08} should be taken into account. Anyway, those simple estimates of solar proton spectra for several recent SEP events give serious evidence that the event of 2012 May 17 was really the first GLE of the solar cycle 24.

The value of $\alpha$ (or $\gamma$) obtained above seems to be typical for shock wave acceleration \citep[e.g.,][]{ell85,zan00}. However, as noted by \citet{ber03}, in both above papers the authors have considered a case of plane wave approximation that does not allow to take into account a finite size of the shock wave and its temporal dependence. Such an approximation is applicable to a bulk of accelerated particle in the vicinity of the shock, but it is broken in the range of ultimate energies where the spectrum undergoes to exponential cutoff (see left panel of Figure 9). In fact, this approach results in significant softening of particle spectrum and decreasing of their maximum energy.

On the other hand, recently \citep{kuw12} derived the spectrum of relativistic solar protons for the GLE71 based on the data from the large Antarctic installation, namely, IceTop Cherenkov detectors. They found that the differential spectral index varies from $\alpha \sim$ 4.3 (pulse phase) to $\alpha \sim$ 4.9 (broad enhance phase). This is consistent with the estimates by \citet{vas11}, who have found that the values of differential spectral index for a delayed component of different GLEs are distributed from 4 up to 6. These authors attributed it to the stochastic acceleration in turbulence plasma, which may be connected to an expanding CME.

\section{Summary and discussion}

In this study we combine a wide range of data sets, specifically \emph{in-situ} particle measurements, ground-based detections of SCRs, and remote-sensing solar observations, to identify the acceleration source(s) for electrons and protons during the first GLE event of the solar cycle 24. The SPR times of the two species are derived and compared with the associated flare and CME phenomena. Detailed data on the solar eruptions are investigated and related to the possible particles acceleration source(s) by analyzing event timing, particle energy spectrum, etc.

Data analysis leads to the following results and main conclusions: (1) The SPR time of near-relativistic electrons is derived to be 01:29 $\pm$ 00:01 UT and the IMF path length 1.25 $\pm$ 0.05 AU. (2) The impulsive injection phase of electrons, as indicated by the dramatic change of spectral index, is consistent with the flare non-thermal emission and the type III radio bursts. (3) The PFSS modeled open field lines rooted in the AR provide escaping channels for the flare accelerated electrons. (4) Whereas, the GLE protons injection takes place at 01:39 $\pm$ 00:02 UT, which is $\sim$10 min later than the electrons injection. (5) The GLE proton injection time is in accordance with the type II radio burst and the prominence eruption, which drives a high speed CME. (6) the GLE protons, with an estimated kinetic energy of $\sim$1.12 GeV, are probably accelerated by the CME-driven shock when it travels to $\sim$3.07 Rs.

Those preliminary results imply that, in general, our findings cannot still give certain self-consistent scenario of the GLE71. In particular, more accurate estimates of spectral properties of SEPs are needed, with taking into account interplanetary propagation of accelerated particles. Also, it would be timely to note that up to now the present solar cycle develops rather limply. Faint proton emissivity of the Sun in the first years of the cycle 24 may be evidence of specific nature of this cycle which, most likely, is a turning-point in the course of solar activity during last 150 -- 200 years. In this context, it should be emphasized the importance of adequate interpretation of SEP event data of this cycle, especially on this unique GLE71 event of 2012 May 17.

To complete this discussion it is worth to note that some peculiarities of the oscillations of GLE occurrence rate for the entire period of ground-based observations (since 1942), in principle, had allowed to predict the moment of GLE71 occurrence. Tentative forecast by \citet{per11} indicated that after characteristic 5-year pause (after GLE70 on 2006 December 13) the next GLE to be occurred between 2011 December 12 and 2012 February 2. In fact, the GLE71 occurred on 17 May 2012, i.e. several months after predicted date.

\acknowledgments

The authors are very grateful to the $WIND$, $ACE$, $GOES$, $SDO$, $SOHO$, and $RHESSI$ spacecraft teams, as well as to NMDB community, for their open data policy. This work has been supported by the Natural Science Foundation (BK2012299) of Jiangsu province, the Ph.D. programs foundation of Ministry of Education of China (20120091120034) and the Technology Foundation for Selected Overseas Chinese Scholar, Ministry of Personnel of China. Leonty Miroshnichenko sincerely thanks Prof. W.Q. Gan and his colleagues from the Purple Mountain Observatory (Nanjing, China) for their kind hospitality during his visit to Nanjing in November 2012.

\clearpage

\begin{figure}
\epsscale{.70}
\plotone{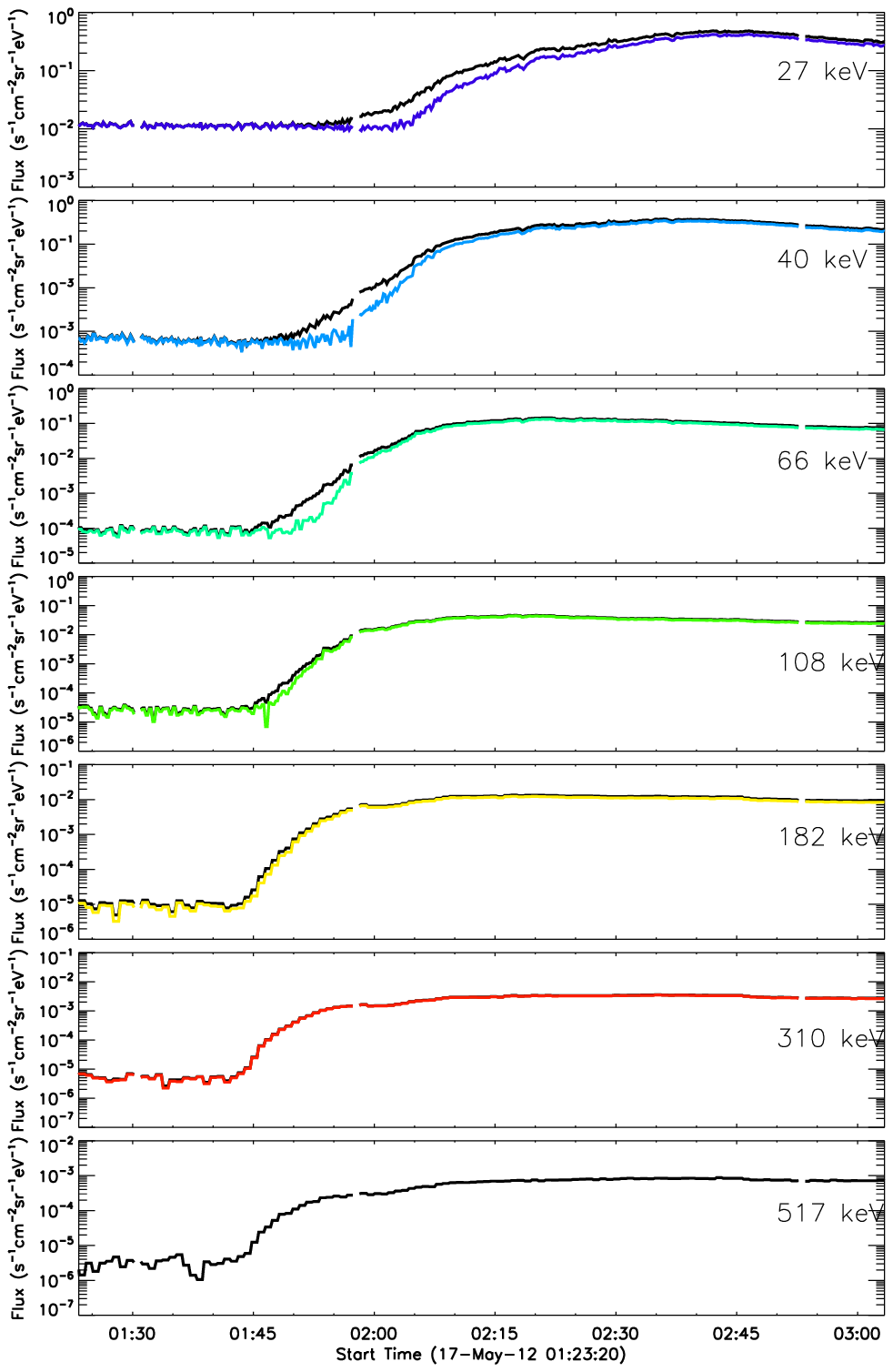}
\caption{Electron intensity profiles detected by $WIND$/3DP/SSTs in seven channels on 2012 May 17. Black lines indicate the original fluxes, and color lines show fluxes after correction for scatter-out electrons.\label{fig1}}
\end{figure}

\begin{figure}
\epsscale{.70}
\plotone{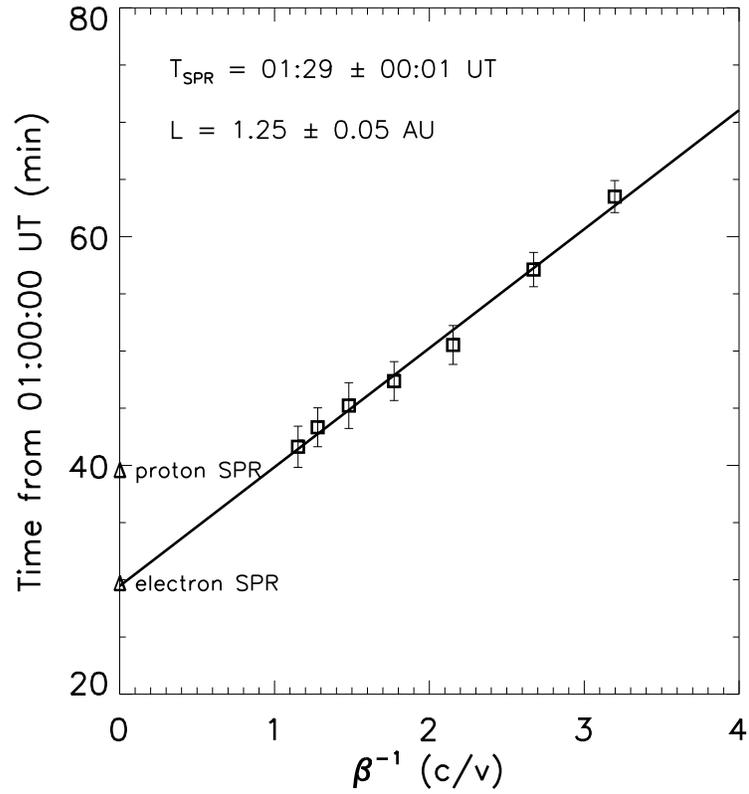}
\caption{Onset times of electron fluxes in seven channels as a function of inverse velocity. The straight line is the linear regression to the observations. The interception gives the SPR time of electrons and the slope corresponds to the IMF path length. The SPR times of electrons and protons are marked by triangles on the time axis.\label{fig2}}
\end{figure}

\begin{figure}
\epsscale{.70}
\plotone{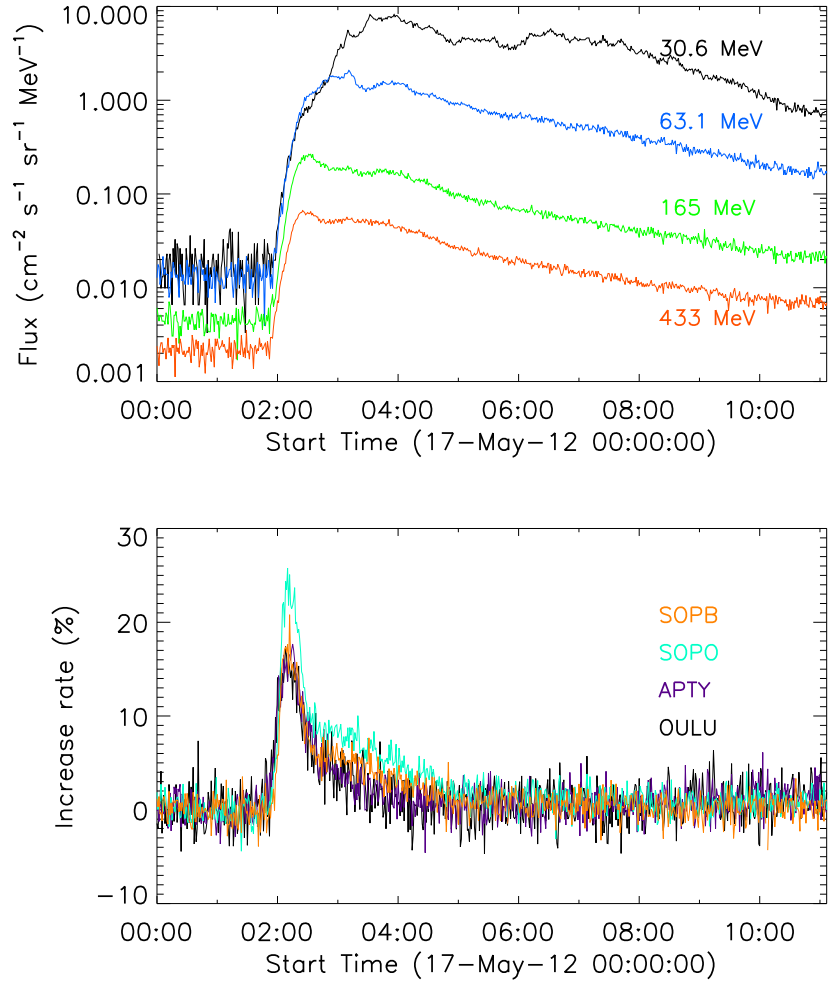}
\caption{GLE event on 2012 May 17. Differential proton intensity profiles detected by the $GOES$ 13 in four energy channels (top panel) and GLE of SCR increases recorded by four NMs (bottom panel).\label{fig3}}
\end{figure}

\begin{figure}
\epsscale{.60}
\plotone{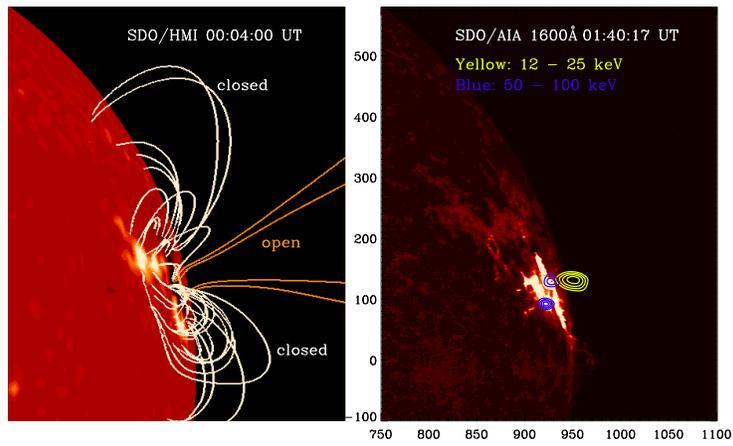}
\caption{PFSS modeled coronal magnetic configurations of the AR 11476 (left panel) and two-ribbon flare in $SDO$/AIA 1600 $\rm {\AA}$ (right panel), overlaid with the $RHESSI$ HXR sources. Blue contour lines indicate 50--100 keV HXR source, and yellow 12--25 keV, integrated from 01:41:30 UT through 01:42:30 UT.\label{fig4}}
\end{figure}

\begin{figure}
\epsscale{.70}
\plotone{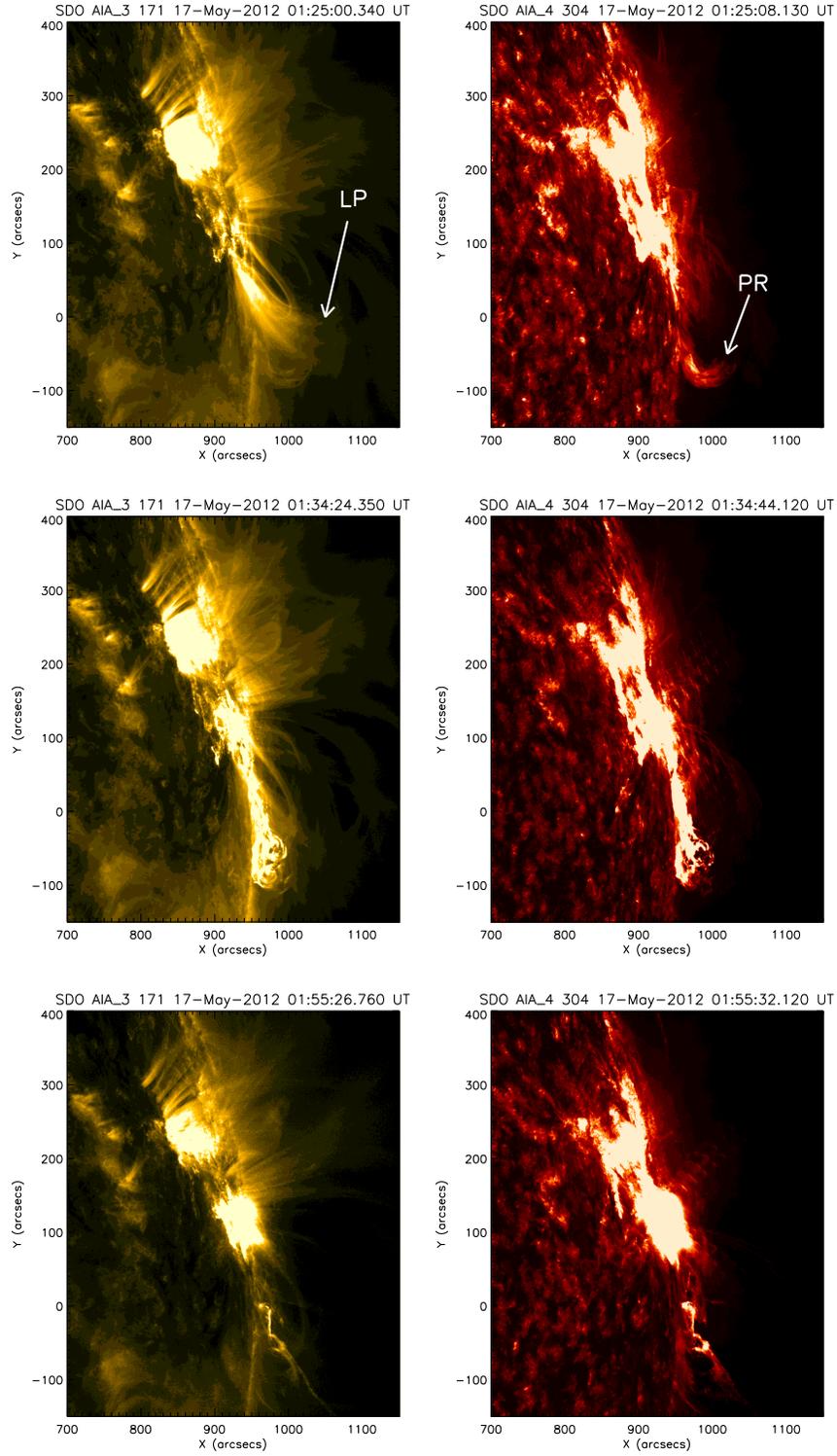}
\caption{Temporal evolution of the AR 11476 in $SDO$/AIA 171 $\rm {\AA}$ and 304 $\rm {\AA}$. The labels ``LP" and ``PR" indicate the loop system and prominence, respectively.\label{fig5}}
\end{figure}

\begin{figure}
\epsscale{.60}
\plotone{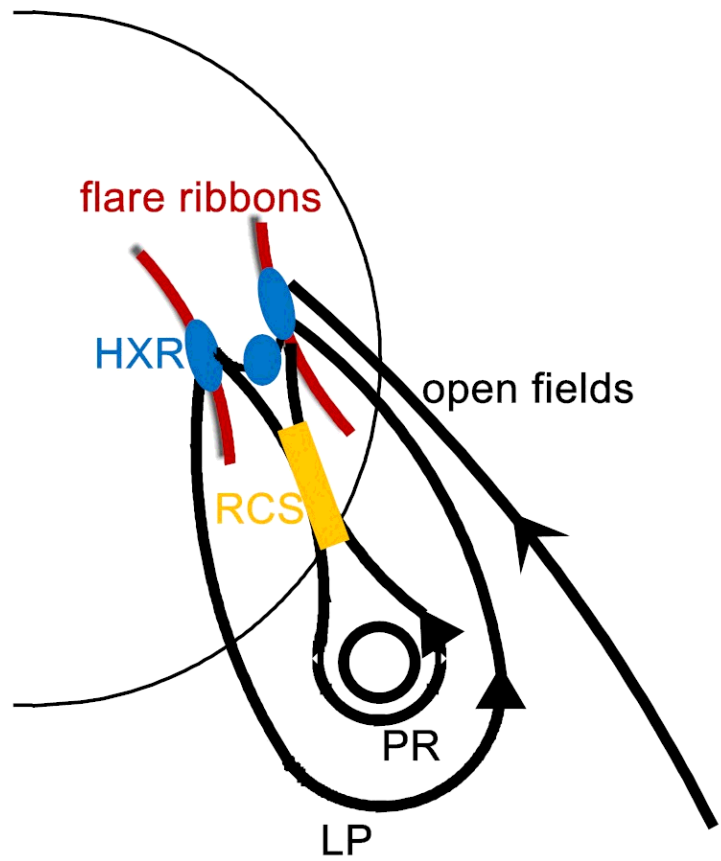}
\caption{A sketch of the magnetic topology associated with the two-ribbon flare and prominence eruption based on Figure 4 and 5.\label{fig6}}
\end{figure}

\begin{figure}
\epsscale{.60}
\plotone{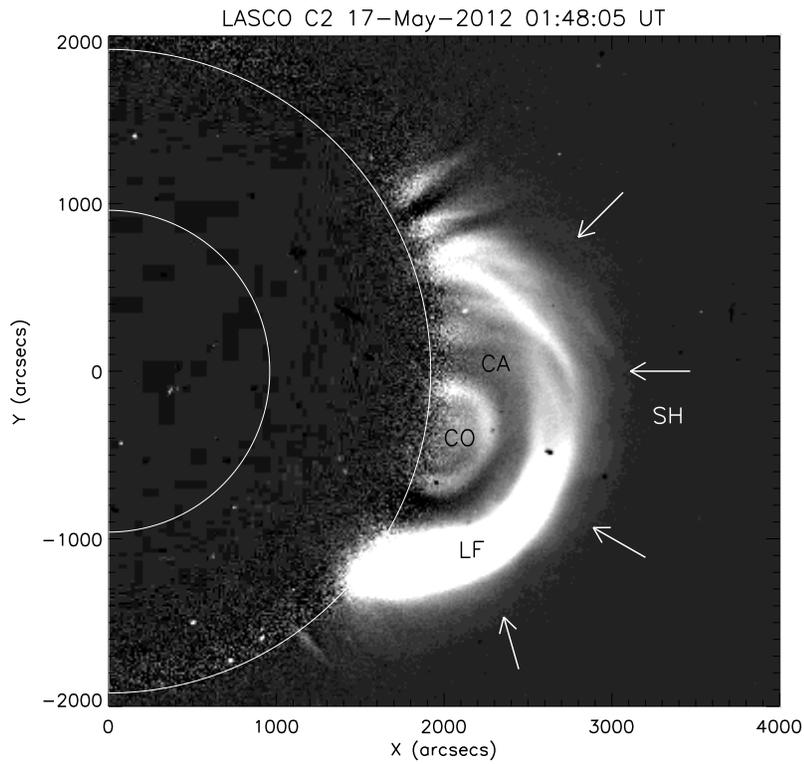}
\caption{$SOHO$/LASCO C2 running difference image with a previous image subtracted. The labels ``CO", ``CA", and ``LF" indicate the three CME components: core, cavity, and leading front, respectively. ``SH" indicates the CME-driven shock structure.\label{fig7}}
\end{figure}

\begin{figure}
\epsscale{.70}
\plotone{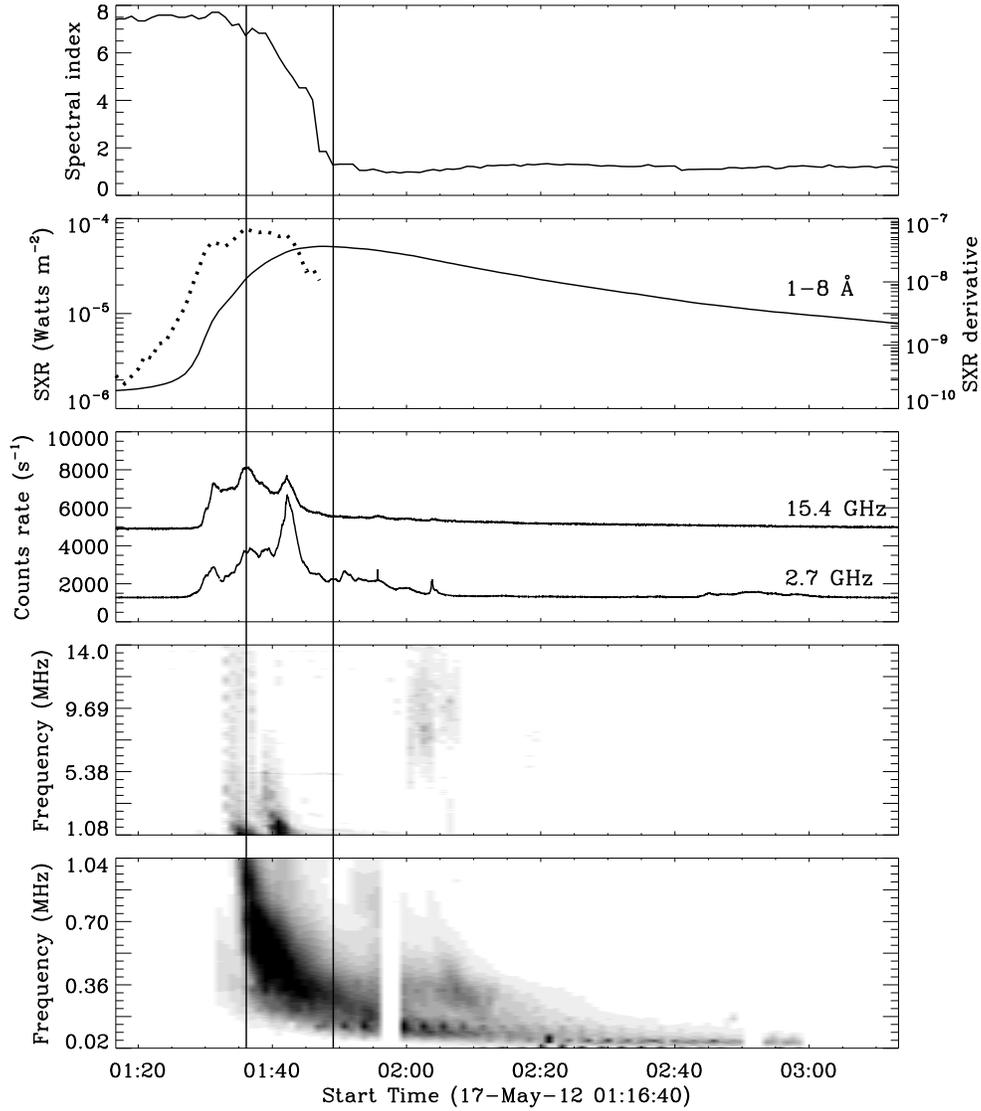}
\caption{From the top to the bottom: evolution of electron spectral index, $GOES$ SXR flux in 1--8 $\rm {\AA}$ and its derivative in positive values (dotted curve), RSTN/Learmonth microwave emission in 15.4 GHz and 2.7 GHz, and the $WIND$/WAVES radio dynamic spectra in frequency range of 20 kHz to 14 MHz. The time range between two vertical lines indicate the electron impulsive injection phase.\label{fig8}}
\end{figure}

\begin{figure}
\epsscale{.70}
\plotone{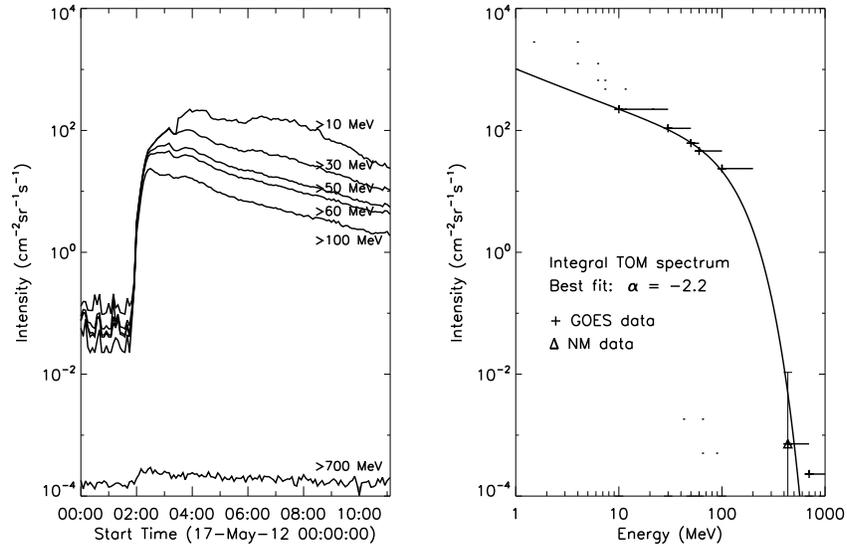}
\caption{Left panel: integral proton intensity profiles measured in six energy channels onboard $GOES$ 13 on 17 May 2012. Bottom panel: TOM power-law spectrum within energy range of $>$10 -- $>$100 MeV truncated by an exponential high-energy tail.\label{fig9}}
\end{figure}

\clearpage

\begin{deluxetable}{lccccc}


\tablecolumns{6} \tablewidth{0pc} \tablecaption{Time-of-maximum spectrum for protons
in several major SEP events} \tablehead{
\colhead{Event} & \colhead{$I_{(>10\rm MeV)}$} & \colhead{$I_{(>50\rm MeV)}$} &
\colhead{$I_{(>100\rm MeV)}$}  & \colhead{spectral index} & \colhead{$I_{(>433\rm MeV)}^{\ast}$}\\
\colhead{} & \colhead{($pfu$)} & \colhead{($pfu$)} &
\colhead{($pfu$)}  & \colhead{$\alpha$} & \colhead{($pfu$)}}

\startdata
1972 Aug 04 & $1.0 \times 10^{5}$ & $2.0 \times 10^{4}$ & $1.0 \times 10^{3}$ & 1.0 & $3.0 \times 10^{0}$ \\
2012 Jan 27 & 796 & 32 & 8 & 2.0 & $n/a$ \\
2012 Mar 07 & 6530 & 305 & 70.5 & 2.0 & $n/a$ \\
2012 May 17 & 222 & 62 & 23 & 2.2 & $7.1 \times 10^{-4}$ \\
\enddata

\tablecomments{$I$ is the integral intensity of solar protons by $GOES$ 13 data. $I^{\ast}$ is derived by real NM data. $n/a$ indicates no NM data available.}

\end{deluxetable}

\end{document}